\documentclass[11pt, titlepage, reqno]{amsart}% The first page is devoted to the title and eqs numbers appear on the right
\usepackage[utf8]{inputenc}
\usepackage[T1]{fontenc}
\usepackage[british]{babel}
\usepackage{times}
\usepackage{amsaddr, amsbsy, amscd, amsfonts, amsmath, mathrsfs, amssymb, amsthm}
\usepackage{dsfont, mathrsfs, mathtools}
\usepackage{hyperref}
\usepackage{enumitem}
\usepackage{xparse}
\usepackage[bottom, hang, flushmargin]{footmisc}

\usepackage[a4paper, hmargin=2cm, vmargin={1.5cm,2.5cm}, nomarginpar, hcentering, includeheadfoot]{geometry}
\usepackage{eqnarray}
\usepackage{scalerel, stackengine}
\usepackage{enumitem}
\usepackage{fancyhdr}
\usepackage{indentfirst}

\usepackage[toc,page]{appendix}
\theoremstyle{definition}
\newtheorem{defn}{Definition}

\theoremstyle{plain}

\newtheorem{theo}{Theorem}[section]

\newtheorem{prop}[theo]{Proposition}

\numberwithin{equation}{section}
\numberwithin{defn}{section}
\numberwithin{note}{section}

\stackMath

\let\oldker\ker
\let\oldforall\forall
\let\oldexists\exists
\renewcommand\widehat[1]{%
\savestack{\tmpbox}{\stretchto{%
    \scaleto{%
        \scalerel*[\widthof{\ensuremath{#1}}]{\kern.1pt\mathchar"0362\kern.1pt}%
        {\rule{0ex}{\textheight}}%WIDTH-LIMITED CIRCUMFLEX
    }{\textheight}% 
}{2.4ex}}%
\stackon[-6.9pt]{#1}{\tmpbox}%
}

\NewDocumentCommand \fromToParser {m m m}{%
    \IfValueTF{#3}{%
        \fromTo{#1}{#2}[#3]%
    }{%
        \fromTo{#1}{#2}%
    }%
}
\NewDocumentCommand \fromTo {m m o}{%
    #1\! \IfValueTF{#3}{%
            \to[#3]
        }{%
            \to[\!\quad\!]
        } \!#2%
}
\newcommand{\integrand}[2]{\!#2\; #1}
\NewDocumentCommand \integralLimits {m m}{%
    _{#1}\IfValueT{#2}{^{\,#2}\!}%
}
\newcommand{\defineSymbol}[2]{#1 \vcentcolon = #2}
\newcommand{\enifedSymbol}[2]{#1 = \vcentcolon #2}

\RenewDocumentCommand \newline {o}{%
\hfill\\[\IfValueTF{#1}{#1}{0} pt]
}
\renewcommand{\-}{\mspace{-1.5mu}}
\newcommand{\+}{\mspace{1.5mu}}
\newcommand{\nquad}{\mspace{-18mu}}

\newcommand{\n}{\noindent}
\newcommand{\vs}{\vspace{0.5cm}}

\providecommand{\comment}[1]{}
\newcommand{\say}[1]{``#1''}

\newcommand{\pInfty}{ {\scriptstyle +}\+\infty}

\newcommand{\Id}{\mathds{1}}
\renewcommand{\forall}{\oldforall\,}
\RenewDocumentCommand \exists {s}{%
    \IfBooleanTF{#1}{\oldexists!}{\oldexists} \;%
}
\ProvideDocumentCommand \define {s >{\SplitArgument{1}{;}} m}{%
    \IfBooleanTF{#1}{\enifedSymbol #2}{\defineSymbol #2}%
}
\RenewDocumentCommand \to {o}{%
    \IfValueTF{#1}{%
        \xrightarrow[\,#1\,]{\;}%
    }{%
        \,\rightarrow\,%
    }%
}
\NewDocumentCommand \maps {m >{\SplitArgument{2}{;}} m}{%
    #1\!: \fromToParser #2%
}
\ProvideDocumentCommand \conjugate {s m}{%
    \IfBooleanTF{#1}{%
        \overline{#2}%
    }{%
        \bar{#2}%
    }%
}
\providecommand{\abs}[1]{\lvert#1\rvert}
\renewcommand*{\vec}[1]{\boldsymbol{#1}}

\NewDocumentCommand \ball {s m m}{%
    \IfBooleanTF{#1}{%
        \mathcal{B}^{\+ c}_{#2}(#3)%
    }{%
        \mathcal{B}_{#2}(#3)%
    }%
}
\NewDocumentCommand \Char { m o }{% Characteristic function 
    \Id_{#1}
    \IfValueT{#2}{( #2 )}
}
\newcommand*{\oSmall}[1]{o\!\left(#1\right)}
\newcommand*{\oBig}[1]{\mathcal{O}\!\left(#1\right)}
\NewDocumentCommand \integrate { >{\SplitArgument{1}{;}} o >{\SplitArgument{1}{;}} m}{%
    \int \IfValueT{#1}{\integralLimits #1 \!} \integrand #2%
}

\NewDocumentCommand \hilbert{s}{%
    \IfBooleanTF{#1}{\mathfrak{H}}{\mathscr{H}}%
}
\NewDocumentCommand \X{s}{%
    \IfBooleanTF{#1}{\mathcal{X}}{\mathfrak{X}}%
}
\NewDocumentCommand \scalar { m m o }{%
    \langle#1,\,#2\rangle\IfValueT{#3}{_{#3}}%
}
\NewDocumentCommand \norm { m o }{%
    \left\lVert#1\right\rVert \IfValueT{#2}{_{#2}}%
}
\NewDocumentCommand \normConverge {o m m o}{
    \IfValueTF{#4}{
        \fromTo{ \IfValueTF{#1}{\norm{#2 - #3}[#1]\! }{\norm{#2 - #3}} }{\,0}[#4]
    }{
        \fromTo{ \IfValueTF{#1}{\norm{#2 - #3}[#1]\! }{\norm{#2 - #3}} }{\,0}
    }
}
\NewDocumentCommand \weakConverge {m m o}{%
    #1 \!\IfValueTF{#3}{%
            \xrightharpoonup[#3]{\quad}
        }{%
            \xrightharpoonup{\quad}%
        } \- #2
}

\newcommand*{\dom}[1]{\mathscr{D}(#1)}
\newcommand*{\ran}[1]{\mathrm{ran}(#1)}
\renewcommand*{\ker}[1]{\oldker(#1)}

\newcommand*{\bounded}[1]{\mathscr{B}\mspace{-1mu}\left(#1\right)}
\NewDocumentCommand \resolvent {m o}{%
    \mathcal{R}_{#1}\IfValueT{#2}{(#2)}%
}
\NewDocumentCommand \spectrum {o m}{%
    \IfValueTF{#1}{%
        \sigma_{\mathrm{#1}}(#2)%
    }{%
        \sigma(#2)%
    }%
}

\NewDocumentCommand \Lp {s m o} { L^{\- #2}\IfBooleanT{#1}{_{\,\mathrm{loc}}}\IfValueT{#3}{(#3)} }
\NewDocumentCommand \LpS {s m o}{
\IfBooleanTF{#1}{\Lp{#2}_{+}}{\Lp{#2}_{\mathrm{sym}}}\IfValueT{#3}{(#3)}
}
\NewDocumentCommand \LpA {s m o}{
\IfBooleanTF{#1}{\Lp{#2}_{-}}{\Lp{#2}_{\mathrm{asym}}}\IfValueT{#3}{(#3)}
}
\NewDocumentCommand \LpSA {m o}{
\Lp{#1}_{\pm}\IfValueT{#2}{(#2)}
}

\NewDocumentCommand \FT { s m o }{
    \IfBooleanTF{#1}{ \IfValueTF{#3}{ (\mathcal{F} \+  #2)\-(#3) }{%
    \mathcal{F} \+  #2 } }{ \hat{#2}\IfValueT{#3}{(#3)} }
}

\newcommand{\C}{\mathbb{C}}
\newcommand{\R}{\mathbb{R}}
\newcommand{\Rplus}{\mathbb{R}_{+}}

\NewDocumentCommand \Z {o}{%
    \IfValueTF{#1}{%
        \mathbb{Z}_{\,#1}%
    }{%
        \mathbb{Z}%
    }%
}

\title[Quantum Hamiltonians with Contact Interactions]{Hamiltonians for Quantum Systems with Contact Interactions}

\author[D. Ferretti]{Daniele Ferretti}
\address{\emph{Gran Sasso Science Institute}, Via Michele Iacobucci, 2 - 67100 (AQ), Italy\\ \textsf{daniele.ferretti@gssi.it}}

\author[A. Teta]{Alessandro Teta}
\address{\emph{Sapienza Universit\`a di Roma}, Piazzale Aldo Moro, 5 - 00185 (RM), Italy\\ \textsf{teta@mat.uniroma1.it}}
\date{}

\thanks{Finanziato dall'Unione Europea - Next Generation EU.\newline
The authors also acknowledge the support of the GNFM Gruppo Nazionale per la Fisica Matematica - INdAM}

\begin{document}

\makeatletter

\newcommand{\pushright}[1]{\ifmeasuring@#1\else\omit\hfill$\displaystyle#1$\fi\ignorespaces}
\newcommand{\pushleft}[1]{\ifmeasuring@#1\else\omit$\displaystyle#1$\hfill\fi\ignorespaces}

\renewcommand\section{%
  \@startsection{section}%
    {1}% level
    {0em}% indent
    {1.5cm \@plus 0.1ex \@minus -0.05ex}% Beforeskip
    {0.75cm \@plus 0.2em}% Afterskip
    {\centering \large \scshape}% Style
  }

  \renewcommand\subsection{%
  \@startsection{subsection}%
    {2}% level
    {0em}% indent
    {0.75cm \@plus 0.1ex \@minus -0.05ex}% Beforeskip
    {0.25cm \@plus 0.2em}% Afterskip
    {\bf\scshape}% Style
  }

\makeatother

\begin{abstract}
We discuss the problem of constructing self-adjoint and lower bounded Hamiltonians for a system of $n>2$ non-relativistic quantum particles in dimension three with contact (or zero-range or $\delta$) interactions.
Such interactions are described by (singular) boundary conditions satisfied at the coincidence hyperplanes, \emph{i.e.}, when the coordinates of two particles coincide.
Following the line of recent works appeared in the literature, we introduce a boundary condition slightly modified with respect to the usual boundary condition one has in the one-body problem.
With such new boundary condition we can show that the instability property due to the fall to the center phenomenon described by Minlos and Faddeev in 1962 is avoided and then one obtains a physically reasonable Hamiltonian for the system.
We apply the method to the case of a gas of $N$ interacting bosons and to the case of $N$ distinguishable particles of equal mass $M$ interacting with a different particle.
In the latter case we also discuss the limit of the model for $M \longrightarrow \pInfty$.
We show that in the limit one obtains the one-body Hamiltonian for the light particle subject to $N$ (non-local) point interactions placed at fixed positions.
We will verify that such non-local point interactions do not exhibit the ultraviolet pathologies that are present in the case of standard local point
interactions.
\newline[10]
\begin{footnotesize}
 
\n \emph{MSC 2020: 
    81Q10; %    Self-adjoint operator theory in quantum theory, including spectral analysis
    81Q15; %    Quantum theory - Perturbation theories for operators and differential equations
    %47B25; %   Symmetric and self-adjoint operators (unbounded)
    70F07; %    Mechanics of particles and systems - Three-body problems
    46N50; %    Functional analysis - Applications in quantum physics
    %47A99;%    Operator theory - None of the above, but in this section
    %35B25; %   Partial differential equations - Singular perturbations
    81V70;%     Quantum theory - Many-body theory; quantum Hall effect
    %35P15;%    Partial differential equations - Estimation of eigenvalues, upper and lower bounds
    %81Q05.%    Quantum theory - Closed and approximate solutions to the Schrödinger, Dirac, Klein-Gordon and other equations of quantum mechanics
}
\end{footnotesize}
\end{abstract}
\maketitle

%\tableofcontents 
%\rhead[\fancyplain{}{\bfseries\leftmark}]{\fancyplain{}{\bfseries\thepage}}
%\lhead[\fancyplain{}{\bfseries\thepage}]{\fancyplain{}{\bfseries Index}}

%\newpage\null\thispagestyle{empty}

\section{Introduction}
For a system of non relativistic quantum particles at low temperature the typical thermal wavelength associated to the particles is usually much larger than the range of the two-body interaction.
In this situation the system exhibits a universal behavior, \emph{i.e.}, the relevant observables do not depend on the details of the two-body interaction but only on few physical parameters, typically the scattering length

\[
a \vcentcolon= \lim_{\abs{\vec{k}}\to 0} f(\vec{k},\vec{k}')\big|_{\abs{\vec{k}}=\abs{\vec{k}'}} 
\]

\n
where $\;f(\vec{k}, \vec{k}')\;$ is the two-body scattering amplitude.

\n
In this regime it may be reasonable to describe the effective behavior of the system by a Hamiltonian with contact (or zero-range or $\delta$) interactions formally given by

\begin{equation}\label{formal}
  -\sum_{i\+=1}^N \Delta_{\vec{x}_i} + \sum_{i\,<\,j}^N \nu_{ij} \+\delta (\vec{x}_i - \vec{x}_j)  
\end{equation}
where $\nu_{ij}$ is related to the scattering length associated to the scattering process between particles $i$ and $j$.

\n
In the physical literature such Hamiltonians have been often used at formal level in both the case of one-particle systems and many-body systems.
As examples, we can mention the Kronig-Penney model of a crystal~\cite{KP}, the bound state problem for the neutron-proton system~\cite{BP}, the scattering of neutrons from condensed matter~\cite{F}, the Efimov effect in the three-body problem~\cite{E} and the derivation of Lee-Huang-Yang formula for a dilute Bose gas~\cite{LHY}. 

\n
From the mathematical point of view, the first problem is to construct the rigorous counterpart of the formal Hamiltonian with contact interactions as a self-adjoint (s.a.) and, possibly, lower bounded operator in a suitable Hilbert space. 

\n
The aim of this note is to review some mathematical results recently obtained in this direction (\cite{BCFT, FFT, FeT2, FeT,  FeT3, FST, FiT, FiT2}, for a different approach see also~\cite{GM, Miche}).

\n
We start with a precise mathematical definition of a Hamiltonian with contact interaction.
Let us consider an interaction supported by a (sufficiently regular) set of Lebesgue measure zero $\Sigma \subset \R^s, s\!\geq \!1$, with codimension at most three, and let us introduce in $L^2(\R^s)$ the symmetric, but not s.a., operator 
\[
\dot{\mathcal{H}}_0 \-=\- -\Delta, \qquad \dom{\dot{\mathcal{H}}_0} = \{ u \in H^2(\R^s) \; \;\text{s.t.} \;\; u\big|_\Sigma =0 \} \,.
\]

\n
Then a natural definition is the following

\begin{defn} \label{def1}
The Schr\"odinger operator $\mathcal{H}\!=\!-\Delta \++\, \delta_{\Sigma}\+$ in $\,L^2(\R^s)$  is a non trivial s.a. extension of $\,\dot{\mathcal{H}}_0, \dom{\dot{\mathcal{H}}_0}$.
\end{defn}

\n
Notice that if the codimension of $\Sigma$ is larger than three one can show that $\mathcal{H}, \dom{\mathcal{H}}$ coincides with  the free Laplacian.
Roughly speaking, one may think of $\mathcal{H}, \dom{\mathcal{H}}$ as a s.a. operator characterized by the two conditions: 
\begin{enumerate}[label=(\textit{\alph*}), itemsep=0.2cm]

    \item $\mathcal{H} \psi = -\Delta \psi\,$ for $\,\psi \in \dom{\mathcal{H}}\,$ s.t. $\,\psi\big|_{\Sigma}=0;$
    \item $\psi \in \dom{\mathcal{H}}$ satisfies some (singular) boundary condition (b.c.) on $\Sigma.$
\end{enumerate}

\n
The main mathematical problem is the construction of such s.a. extensions.
According to the general theory (\cite{P1,P2}, see also~\cite{BHS}) it turns out that the s.a. extensions can be parametrized by a s.a. operator $\Theta$ acting in $L^2(\Sigma)$ but, except in few simple cases, the choice of a \say{physically reasonable} $\Theta$ (and correspondingly of a b.c. on $\Sigma$) may be non trivial.

\n
Let us recall the known results in the simple and well understood case of the one-body problem in $\R^d$, $d \geq 1$, with a point interaction placed for simplicity at the origin so that one has $\Sigma =\{\vec{0}\}$ (see~\cite{Albeverio}).  

\n
The deficiency indices of the operator $\dot{\mathcal{H}}_0,\, \dom{\dot{\mathcal{H}}_0}$ are $(1,1)$ in this case  and therefore all the s.a. extensions $\mathfrak{h}_{\alpha,\+\vec{0}}$, $\alpha \in \R$, can be explicitly constructed.
It turns out that the b.c. satisfied by $\psi \in \dom{\mathfrak{h}_{\alpha,\+\vec{0}}}$ depends on the dimension $d$.
In particular one has
\begin{align}
    &\frac{d \psi}{dx}\big|_{x\+=\+0^+} \! - \frac{d \psi}{dx}\big|_{x\+=\+0^-}\- = \alpha \,\psi \big|_{x\+=\+0}\tag{\textit{d}\,=\,1}\\
    &\psi(\vec{x}) =\;  q \log\frac{1}{\abs{\vec{x}}} + \alpha\, q  + \oSmall{1}, \qquad  \text{for } \,\abs{\vec{x}}\to 0, \quad q \in \C\tag{\textit{d}\,=\,2}\\
    &\psi(\vec{x}) =\: \frac{q}{\abs{\vec{x}}} + \alpha \, q + \oSmall{1}, \;\;\;\; \;\;\;\; \qquad  \text{for }  \,\abs{\vec{x}}\to 0, \quad q \in \C\tag{\textit{d}\,=\,3}
\end{align}
In $d=1$ the parameter $\alpha$ is the strength of the delta interaction.
We stress that in $d\in\{2,3\}$ an element of the domain is singular at the origin, with the same singularity of the Green's function, and the b.c. consists in the fact that the constant term in the asymptotic expansion is proportional to the coefficient of the singular term. We also observe that the parameter $-\alpha^{-1} $ has the physical meaning of scattering length associated to the interaction.

\n
Let us consider the $N\-$-body problem, \emph{i.e.}, the problem to construct~\eqref{formal} as a s.a. operator in $\Lp{2}[\R^{Nd}]$, with $N\!\geq\- 3$ and $d \-\geq \!1$.
Now the set $\Sigma$ is the union of the coincidence hyperplanes $\pi_{ij}\!\vcentcolon=\- \left\{(\vec{x}_1,\ldots,\vec{x}_N)\!\in\-\R^{dN} \,\big| \; \vec{x}_i\-=\vec{x}_j\right\}$, where $i,j\-\in\-\{1, \ldots, N\}$, $i \neq j$, and $\vec{x}_i$ denotes the coordinate of the $i$-th particle in $\R^d$.
In this case the deficiency indices of $\dot{\mathcal{H}}_0, \, \dom{\dot{\mathcal{H}}_0}$ are infinite and it is not a priori clear how to chose a specific b.c. on each hyperplane in order to have a physically reasonable s.a. extension. 

\n
In this situation it seems natural to proceed by analogy with the one-body problem. 

\n
For $d\-=\!1$ one can define an operator 
$\mathcal{H}_{\underline{\alpha}}$, $\underline{\alpha} =(\alpha_{ij})^N_{i\+<\+j}\+,$ with $\; \alpha_{ij}\in \R$, in $\Lp{2}[\R^N]$ satisfying condition $(a)$ and the following b.c. on each hyperplane $\pi_{ij}$

\begin{equation}\left( \frac{\partial\!\-}{\partial x_j\!\-} - \frac{\partial\!}{\partial x_i\!}\right)\- \psi \big|_{x_j=\+x_i^+ } \- - \left( \frac{\partial\!\-}{\partial x_j\!\-} - \frac{\partial\!}{\partial x_i\!}\right)\- \psi \big|_{x_j=\+x_i^- } = \alpha_{ij} \,\psi \big|_{x_j=\+x_i}.
\end{equation}

\n
Using perturbation theory of quadratic forms, it is not difficult to prove that the operator is s.a. and bounded from below. Notice that the above b.c. is the direct generalization of the b.c. in the one-body case.
We stress that we have considered a genuine two-body b.c. since the parameters $\alpha_{ij}$ do not depend on the positions of the other particles $\vec x_k$, $k \neq i,j$.
Moreover, the Hamiltonian $\mathcal{H}_{\underline{\alpha}}$ can be obtained as the norm resolvent limit of Hamiltonians with smooth, rescaled, two-body potentials~(\cite{BCFT1}, \cite{GHL}).

\n
For $d\in\-\{2,3\}$, the codimension of $\pi_{ij}$ is larger than $1$ and therefore perturbation theory does not work. 
For $d\!=\!2$ one can proceed defining the relevant extensions of $\dot{\mathcal{H}}_0, \, \dom{\dot{\mathcal{H}}_0}$ as the operators satisfying condition $(a)$ and the b.c. on $\pi_{ij}$ 
\begin{equation}\label{stmBC2}
    \psi(\vec{x}_1,\ldots,\vec{x}_N)=\;\xi_{ij} \!\left(\tfrac{m_i \vec{x}_i+\+ m_j \vec{x}_j}{m_i + m_j }, \vec{X}_{\!ij} \!\right)\log \frac{1}{\abs{\vec{x}_i-\+\vec{x}_j}}\++\alpha_{ij}\,\xi_{ij} \!\left(\tfrac{m_i \vec{x}_i+\+ m_j \vec{x}_j}{m_i + m_j }, \vec{X}_{\!ij}  \!\right)\-+\oSmall{1}\-,
\end{equation}
for some function $\maps{\xi_{ij}}{\pi_{ij};\C}$, $\alpha_{ij}\!\in\-\R$ and the shortcut $\vec{X}_{\!ij}\!=\-(\vec{x}_1,\ldots,\vec{x}_{i-1},\vec{x}_{i+1},\ldots,\vec{x}_{j-1},\vec{x}_{j+1},\ldots,\vec{x}_N\-)$ (in case $i\!<\!j$).
Using a suitably renormalized quadratic form one can prove that such operator is s.a. and bounded from below (\cite{DFT}, see also~\cite{DR}).
Also in this case the b.c. on $\pi_{ij}$ generalizes that of the one-body problem and it is a genuine two-body b.c. since $\alpha_{ij}$ does not depend on the positions of the other particles $\vec{x}_k,\+k \notin\-\{i,j\}$.

\n
It has also been proven that the Hamiltonian is the norm resolvent limit of approximating Hamiltonians with smooth, two-body, rescaled potentials and with a suitable renormalization of the coupling constant~\cite{GH}.

\n
For $d\!=\!3$ the problem is more subtle and proceeding by analogy with the one-body case one arrives at a result which is unsatisfactory from the physical point of view (we just recall that the situation is rather different for systems made of two species of fermions, see, \emph{e.g.},~\cite{CDFMT, CDFMT1, FT} and~\cite{MS, MS1}).

\n
Indeed, let us consider a system of $N\-\!\geq\! 3$ identical spinless bosons of mass $\frac{1}{2}$ interacting with each other via contact or zero-range interactions.
The Hilbert space of the system is therefore given by
\begin{equation}
    \hilbert_N\vcentcolon=\LpS{2}[\R^{3N}].
\end{equation}

\n
Following the analogy with the one-body problem, one defines an operator $\mathcal{H}_\alpha$ in $\hilbert_N$ satisfying condition $(a)$ and the b.c. on $\pi_{ij}$
\begin{equation}\label{stmBC}
    \psi(\vec{x}_1,\ldots,\vec{x}_N)=\;\frac{\xi\!\left(\frac{\vec{x}_i+\+\vec{x}_j}{2},\vec{X}_{\!ij}\!\right)\!}{\abs{\vec{x}_i-\vec{x}_j}}\,+\-\alpha\,\xi\!\left(\tfrac{\vec{x}_i+\+\vec{x}_j}{2},\vec{X}_{\!ij}\!\right)\-+\oSmall{1}\-,
\end{equation}
with $\maps{\xi}{\pi_{ij};\C}$ and $\alpha \-\in \-\R$.
We stress once again that~\eqref{stmBC} is a two-body boundary condition on $\pi_{ij}$ since $\alpha$ does not depend on the positions of the other particles $\vec{x}_k,\+k \notin\-\{i,j\}$. 
The operator $\mathcal{H}_{\alpha}$ defined in this way is known as Ter-Martirosyan Skornyakov extension.
We point out that the parameter $\alpha$ is related to the two-body scattering length $a$ via the relation
\begin{equation}\label{2BodyScatteringLenght}
   a = -\frac{1}{\+\alpha\+}\+.
\end{equation}

\n
As a matter of fact, already in the case $N\!=\-3$ the Ter-Martirosyan Skornyakov extension is symmetric but not s.a.. More specifically, in $1962$ Minlos and Faddeev (\cite{MF}, see also~\cite{MF2}) have shown that the deficiency indices of $\mathcal{H}_\alpha$ are $(1,1)$, then its s.a. extensions are parametrized by a real constant $\beta$ which characterizes the behaviour of the wave function close to the triple coincidence point, \emph{i.e.}, where the positions of the three particles coincide. 
Furthermore for any value of $\beta$ the s.a. extension $\mathcal{H}_{\alpha,\beta}$ is unbounded from below. 
Such instability is essentially due to the fact that the interaction becomes too strong when the three particles are close to each other and this determines a collapse (or fall to the center) phenomenon. 

\n
The result in~\cite{MF} can be interpreted as a no-go theorem.
Indeed, it states that in dimension three for $N\!=\-3$, and a fortiori for $N\!>\-3$, one cannot define a Hamiltonian with contact interactions via the two-body boundary condition~\eqref{stmBC}, where the strength of the interaction between two particles is a constant and therefore it is a genuine two-body interaction.
Conversely, one is forced to add a further three-body boundary condition, corresponding to a sort of three-body force acting between the particles when they are close to each other.
Furthermore, the fact that the s.a. Hamiltonian $\mathcal{H}_{\alpha,\beta} $ constructed in~\cite{MF} is not bounded from below, and then it is unsatisfactory from the physical point of view, simply means that the Ter-Martirosyan Skornyakov extension characterized by~\eqref{stmBC} is not a good starting point to construct the Hamiltonian. 

\n
Following a suggestion given in~\cite{MF} (see also~\cite{AHKW}), we introduce a slightly modified b.c. obtained by replacing the constant $\alpha$ in~\eqref{stmBC} by a function $A$ as follows
\begin{equation}\label{mfBC}
    \psi(\vec{x}_1,\ldots,\vec{x}_N)=\;\frac{\xi\!\left(\frac{\vec{x}_i+\+\vec{x}_j}{2},\vec{X}_{\!ij}\!\right)\!}{\abs{\vec{x}_i-\vec{x}_j}}\++ (A \+\xi)\!\left(\tfrac{\vec{x}_i+\+\vec{x}_j}{2},\vec{X}_{\!ij}\!\right)\-+\oSmall{1}\-,\quad \abs{\vec{x}_i\--\vec{x}_j}\longrightarrow 0,
\end{equation}
where the function $A$ is given by 
\begin{equation}\label{alfa}
    \begin{split}
    &A(\vec{z},\vec{y}_1,\ldots,\mspace{2.25mu}\vec{y}_{N-2}) =  \,\alpha+\gamma\-\sum_{k\+=\+1}^{N-2}\frac{\theta(\abs{\vec{y}_k\--\vec{z}})}{\abs{\vec{y}_k\--\vec{z}}}+\frac{\gamma}{2}\-\sum_{k\,<\,\ell}^{N\--2}\frac{\theta(\abs{\vec{y}_k\--\vec{y}_\ell})}{\abs{\vec{y}_k\--\vec{y}_\ell}}
    %\\=&\,\alpha_0+\frac{\gamma}{2}\!\-\sum_{\substack{1\+\leq\, k\+<\+\ell\,\leq\+ N\\ k\+\leq \+ N-2}} \!\- \frac{\theta(\abs{\vec{y}_k-\vec{y}_\ell})}{\abs{\vec{y}_k-\vec{y}_\ell}}\Big|_{\vec{y}_{N-1}=\,\vec{y}_N=\,\vec{z}}
    \end{split}
\end{equation}
with $\gamma>0$ and $\maps{\theta}{\Rplus;\R}$ an essentially bounded function satisfying
\begin{equation}
    \label{positiveBoundedCondition}
    1-\tfrac{r}{b}\leq\theta(r)\leq 1+\tfrac{r}{b},\qquad\text{for almost every } \+r\in\Rplus\+ \text{ and some }\, b>0.
\end{equation}

\n
We observe that the function $\theta$, by assumption~\eqref{positiveBoundedCondition}, is positive almost everywhere in a neighborhood of the origin, and $\theta(r)\longrightarrow 1$ as $r\to 0^+\-.$
Notice that simple choices for this function are $\theta(r)= e^{-r/b}$ or $\theta(r)= \Char{b}(r)$, where $\Char{b}$ is the characteristic function of the ball of radius $b$ centered in the origin.
One can also choose a smooth $\theta$ with an arbitrarily small compact support. 

\n
From the heuristic point of view, the function $A$ introduces an effective two-body scattering length associated to the pair $i,j$ depending on the positions of the other particles.
Such dependence plays the role of a repulsive force that weakens the contact interaction between the particles $i$ and $j$.
In particular, two kind of repulsions are described by~\eqref{alfa}: the first term represents a three-body force that makes the usual two-body point interaction weaker and weaker as a third particle is approaching the common position of the first two interacting particles, while the second term represents a four-body repulsion meant to compensate the singular ultraviolet behaviour associated to the situation in which two other different particles compose an interacting couple $k,l$, with no elements in common with $i,j$.

\n
In this sense the b.c.~\eqref{mfBC} at $\pi_{ij}$ is not a two-body b.c. due to the dependence of the function $A$ on the positions of the other particles.
On the other hand, we stress once again that we can choose an arbitrarily small, compact support for the function $\theta$ so that the usual two-body b.c. is restored as soon as the other bosons are sufficiently far from the singularities.

\n
Our aim is to consider a Hamiltonian in $\hilbert_N$ satisfying condition $(a)$ and the b.c.~\eqref{mfBC} and to show that for $\gamma$ sufficiently large it can be rigorously defined as a s.a. and bounded from below operator in $\hilbert_N$.

\n
More precisely, in the next section we give the rigorous definition of the Hamiltonian, we formulate the result and we give a hint for the proof.
We also establish the connection between our Hamiltonian and the one obtained by Albeverio, H{\o}egh-Krohn and Streit in 1977 using the theory of Dirichlet forms~\cite{AHK}.
In particular we shall see that the latter is obtained as a special case of our Hamiltonian by a suitable choice of $\alpha$, $\gamma$ and $\theta$.

\n
Then we consider the special case of a system of $N$ distinguishable particles of equal mass $M$ interacting via contact interaction with a different particle with mass $m$.
Finally we will show that in the limiting case $M \to \infty$ one obtains the one-body Hamiltonian for the particle with mass $m$ subject to $N$ (non-local) point interactions placed at fixed positions.
We will verify that such non-local point interactions do not exhibit the ultraviolet pathologies that are present in the case of standard local point interactions.

\section{Hamiltonian for \texorpdfstring{$N$}{N} interacting bosons}
In order to characterize  our Hamiltonian it is first convenient to introduce 
the function  $\mathscr{G}^\lambda\xi$ defined as the distributional solution to
\begin{equation}
(\mathcal{H}_0\-+\-\lambda)\+\mathscr{G}^\lambda\xi=8\pi \+\delta_\pi \,\xi\,.
\end{equation}
The explicit form of such function can be given.  
Indeed, let us denote by $G^\lambda$, with $\lambda\->\-0$, the kernel of the operator $(-\Delta+\lambda)^{-1} \-\in\-\bounded{\hilbert_N,\dom{\mathcal{H}_0}}$ given by 
\begin{equation}\label{greenLaplace}
    G^\lambda\Big(\-
        \begin{array}{c}
            \!\vec{x}\!\-\\[-5pt]
            \!\vec{y}\!\-
        \end{array}\!\Big)\-=\tfrac{1}{\,(2\pi)^{3N/2}\!\-}\left(\-\frac{\!\!\sqrt{\lambda}}{\+\abs{\vec{x}-\vec{y}}\+}\-\right)^{\!\frac{3N}{2}-1}\nquad\mathrm{K}_{\frac{3N}{2}-1}\!\left(\-\sqrt{\lambda}\,\abs{\vec{x}-\vec{y}}\right)\!,\qquad \vec{x},\vec{y}\in\R^{3N},
\end{equation}
where $\maps{\mathrm{K}_\nu}{\Rplus;\Rplus}$ is the modified Bessel function of the second kind (also known as Macdonald's function) of order $\nu\geq 0$~\cite[section 8.43]{GR}.
We recall that 
\begin{subequations}\label{macdonaldProperties}
\begin{gather}
    z^\nu\+\mathrm{K}_\nu(z)\,\text{ is decreasing in }\,z\-\in\-\Rplus\+,\label{decreasingMacdonald}\\[-2.5pt]
    \mathrm{K}_\nu(z)=\frac{2^{\nu-1}\Gamma(\nu)}{z^\nu}+\oSmall{z^{-\nu}}\!,\qquad \text{for }\, z\to[\!\quad\!] 0^+\-, \,\nu>0,\label{originMacdonald}\\
    \mathrm{K}_\nu(z)=\sqrt{\-\frac{\pi}{2\+z}}\,e^{-z}\!\left[1+\frac{4\nu^2-1}{8\+z}+\oBig{z^{-2}}\right]\!,\qquad \text{for }\,z\to[\!\quad\!] \pInfty.\label{infinityMacdonald}
\end{gather}
\end{subequations}
Then the explicit representation of $\mathscr{G}^\lambda\xi$ is
\begin{align}\label{potentialDef}
    (\mathscr{G}^\lambda\xi)(\vec{x}_1,\ldots,\vec{x}_N)\vcentcolon&\mspace{-5mu}=8\pi \sum_{i\,<\,j}
    \int_{\R^{3(N-1)}} \mspace{-36mu} d\vec y d \vec Y_{\!ij} \;\xi(\vec{y},\vec{Y}_{\!\!ij})
\,G^\lambda\-
\bigg(\!
        \begin{array}{r c l}
            \!\-\vec{x}_i\+,& \!\!\-\vec{x}_j\+, & \!\!\-\vec{X}_{\!ij}\!\\[-2.5pt]
            \!\-\vec{y},& \!\!\-\vec{y},& \!\!\-\vec{Y}_{\!\!ij}\!
        \end{array}
        \!\!\bigg)\\
    &=\vcentcolon\sum_{i\,<\,j} \,(\mathscr{G}_{ij}^\lambda\mspace{2.25mu}\xi)(\vec{x}_1,\ldots,\vec{x}_N).\nonumber
\end{align}
With a slight abuse of terminology, given $\{i,j\}$, we refer to $\mathscr{G}^\lambda_{ij}\+\xi$ as the \emph{potential} generated by the \emph{charge} $\xi(\vec{x},\vec{X}_{\!ij}\-)$ distributed along $\pi_{ij}$, while $\mathscr{G}^\lambda\xi$ shall consequently be the total potential. 
Clearly, owing to the bosonic symmetry, each charge associated to the coincidence hyperplane $\pi_{ij}$ is equal to any other charge distributed along another hyperplane $\pi_{kl}$. 
We recall that the operator $\mathscr{G}^\lambda$ is injective and its range is contained in $\hilbert_N$.
Additionally, we stress that $\ran{ \mathscr{G}^\lambda} \cap 
H^1(\R^{3N})=\{0\}$.  
Finally, in the Fourier space we have
\begin{equation}\label{potentialFourier}
    (\widehat{\mathscr{G}^\lambda_{ij}\+\xi})(\vec{p}_1,\ldots,\vec{p}_N)=\sqrt{\tfrac{8}{\pi}\+}\+\frac{\FT{\xi}(\vec{p}_i\-+\vec{p}_j,\vec{P}_{\!ij})}{p_1^2+\ldots+p_N^2+\lambda}.
\end{equation}
In our context, the relevant property of the potential is the asymptotic behavior near each hyperplane
\begin{equation} \label{pot}
(\mathscr{G}^\lambda\xi)(\vec{x}_1,\ldots,\vec{x}_N)=\frac{\xi\big(\tfrac{\vec{x}_i+\+\vec{x}_j}{2}\-,\+\vec{X}_{\!ij}\-\big)}{\abs{\vec{x}_i\--\vec{x}_j}}-(\Gamma_{ij}^\lambda\+\xi)\big(\tfrac{\vec{x}_i+\+\vec{x}_j}{2}\-,\+\vec{X}_{\!ij}\big)+\oSmall{1}\-.
\end{equation}
where the operator $\maps{\Gamma_{ij}^\lambda}{\Lp{2}[\pi];\Lp{2}[\pi_{ij}]}$ is given by
\begin{equation}
    \begin{split}
        (\widehat{\Gamma_{ij}^\lambda\+\xi})(\vec{k},\vec{P}_{\!ij})\!=\-&\;\tfrac{1}{\sqrt{2}}\sqrt{\tfrac{1}{2}\+k^2\-+P^2_{\!ij}\-+\lambda\+}\:\FT{\xi}(\vec{k},\vec{P}_{\!ij})\++\\
        &-\-\tfrac{2}{\,\pi^2\!}\!\!{\textstyle\sum\limits_{\ell\,\notin\+\{i,\+j\}}^N}\-\integrate[\R^6]{\delta(\vec{p}_i\!+\-\vec{p}_j\!-\-\vec{k})\,\frac{\FT{\xi}(\vec{p}_i\-+\vec{p}_\ell,\vec{P}_{\!i\ell})%+\FT{\xi}(\vec{p}_j\-+\vec{p}_\ell,\vec{P}_{\!j\ell})
        }{p_i^2+p_j^2+P_{\!ij}^2+\lambda};\!d\vec{p}_id\vec{p}_j}+\\
        &-\-\tfrac{1}{\,\pi^2\!}\nquad\!\!\!{\textstyle\sum\limits_{\substack{\ell\,<\+k\\ \{\ell,\+k\}\+\cap\+\{i,\+j\}=\emptyset}}^N}\!\!\integrate[\R^6]{\delta(\vec{p}_i\!+\-\vec{p}_j\!-\-\vec{k})\,\frac{\FT{\xi}(\vec{p}_k\-+\vec{p}_\ell,\vec{P}_{\!k\ell})}{p_i^2+p_j^2+P_{\!ij}^2+\lambda};\!d\vec{p}_id\vec{p}_j}.
    \end{split}
\end{equation}
Proceeding at heuristic level, let us denote by $\mathcal H'$ our formal Hamiltonian satisfying condition $(a)$ and the b.c. \eqref{mfBC}. 
Using the potential $\mathcal G^{\lambda} \xi$, with $\xi$ sufficiently smooth, we can characterize $\mathcal H'$ in a more convenient form. 
Indeed, a comparison of the asymptotic behavior~\eqref{pot} with~\eqref{mfBC} suggests that an element of the domain of $\mathcal H'$ can be decomposed as
\begin{equation}
    \psi=w_\lambda\-+\+\mathscr{G}^\lambda\+\xi,\qquad w_\lambda\in H^2(\R^{3N})\cap \LpS{2}[\R^{3N}]
\end{equation}
and consequently the b.c.~\eqref{mfBC} can be rewritten as follows
\begin{equation}
    (A+\-\Gamma^\lambda_{ij})\+\xi\+=w_\lambda|_{\pi_{ij}}.
\end{equation}
Moreover, setting
$\,D_\epsilon=\{(\vec{x}_1,\ldots,\vec{x}_N)\in\R^{3N}\:|\; \min\limits_{i\,<\,j}\abs{\vec{x}_i-\vec{x}_j}>\epsilon\},$
one can make use of condition $(a)$ and the definition of $\mathscr{G}^\lambda\xi$ to obtain
\begin{align*}\scalar{\varphi}{(\mathcal{H}'+\lambda)\psi}[\hilbert_N]&=\lim_{\epsilon\to 0}\integrate[D_\epsilon]{\conjugate*{\varphi(\vec{x}_1,\ldots,\vec{x}_N)}\,(\mathcal{H}_0+\lambda)(w_\lambda\-+\mspace{0.75mu}\mathscr{G}^\lambda\xi)(\vec{x}_1,\ldots,\vec{x}_N); d\vec{x}_1\-\cdots d\vec{x}_N}\\
&=\scalar{\varphi}{(\mathcal{H}_0+\lambda) w_\lambda}[\hilbert_N]
\end{align*}
for any $\varphi$ Schwartz function.
Thus, one has
$$(\mathcal{H}'+\lambda)\psi=(\mathcal{H}_0+\lambda)\+ w_\lambda\quad\iff\quad \mathcal{H}'\psi=\mathcal{H}_0\+ w_\lambda-\lambda\+\mathscr{G}^\lambda\xi.$$
At a rigorous level, the first crucial result one needs to prove is the following.
\begin{prop}\label{crucialResultToProve}
 Let us fix $\gamma > \gamma_c\,$, where
 \begin{equation}\label{criticalGammaN}
    \gamma_c\vcentcolon=2-\frac{8\+\sqrt{3}}{\pi(N\!-\-2)\!\left[8+\sqrt{3}\+(N\--3)\-\right]\!\-}\,.
\end{equation}
Then the operators $A+\Gamma_{ij}^\lambda,\: \dom{A+\Gamma_{ij}^\lambda}=H^1(\pi)$ acting between $\Lp{2}[\pi]$ and $\Lp{2}[\pi_{ij}]$ are s.a. and there exists $\lambda_0>0$ s.t. $A+\Gamma_{ij}^\lambda$ are positive for any $\lambda>\lambda_0.$
\end{prop}
\n
This statement can be proved by considering the quadratic form in $\Lp{2}[\pi]$
\begin{equation} \Phi^\lambda[\xi]=\textstyle\sum\limits_{i\+<\+j}\scalar{\xi}{(A+\Gamma_{ij}^\lambda)\+\xi}[\Lp{2}[\pi_{ij}]],\qquad\dom{\Phi^\lambda}=H^{\frac{1}{2}}(\pi)
\end{equation}
and by showing that $\Phi^\lambda$ is closed for any $\lambda\->\-0$ and bounded from below by a positive constant when $\lambda\->\-\lambda_0$ (for the proof see~\cite{FeT3}). 
Once  proposition~\ref{crucialResultToProve} is proven, the remaining steps easily follow. Indeed, we define the rigorous counterpart $\mathcal{H}_{\-A}$ of the formal Hamiltonian $\mathcal H'$ as follows.
\begin{defn}
    \begin{equation*}
        \begin{dcases}
            \dom{\mathcal{H}_{\-A}}=\Big\{\psi\in\hilbert_N\:\big|\;\psi-\mathscr{G}^\lambda\xi=w_\lambda\!\in\!H^2(\R^{3N}),\, \xi\-\in\! H^1(\pi),\:(A+\-\Gamma_{ij}^\lambda)\+\xi=w^\lambda|_{\pi_{ij}},\,\lambda\->\-0\Big\}\\
            \mathcal{H}_{\-A}\psi=\mathcal{H}_0 \+ w_\lambda-\lambda\+\mathscr{G}^\lambda\xi.
        \end{dcases}
    \end{equation*}
\end{defn}
\n
We stress that this definition does not depend on $\lambda\->\-0$ since it is an arbitrary parameter introduced in the decomposition of $\psi\in\dom{\mathcal{H}_{\-A}}$.

\n
Notice that condition $(a)$ is satisfied, since $\psi\-\in\- H^2_0(\R^{3N}\!\smallsetminus\pi)$ is in $\dom{\mathcal{H}_{\-A}}$ by picking $\xi=0$ and there holds $\mathcal{H}_{\-A}\psi=\mathcal{H}_0\+\psi.$

\n
Additionally, let us  show that $\mathcal{H}_{\-A}$ is a symmetric operator.
Let $\varphi,\psi\in\dom{\mathcal{H}_{\-A}}\,$ with $\,\varphi=v_\lambda+\mathscr{G}^\lambda\eta\,$ and $\,\psi=w_\lambda+\mathscr{G}^\lambda\xi\,$, so that for any $\lambda\->\-0$
\begin{align*}
    \scalar{\varphi}{(\mathcal{H}_{\-A}+\lambda)\+\psi}[\hilbert_N]\!&=\!\scalar{v_\lambda+\mathscr{G}^\lambda\eta}{(\mathcal{H}_0+\lambda)\+w_\lambda}[\hilbert_N]\\
    &=\!\scalar{(\mathcal{H}_0+\lambda)\+v_\lambda}{w_\lambda}[\hilbert_N]\-+\scalar{\mathscr{G}^\lambda\eta}{(\mathcal{H}_0+\lambda)\+w_\lambda}[\hilbert_N]\\
    &=\!\scalar{(\mathcal{H}_0+\lambda)\+v_\lambda}{\psi}[\hilbert_N]\!-\scalar{(\mathcal{H}_0+\lambda)\+v_\lambda}{\mathscr{G}^\lambda\xi}[\hilbert_N]\-+\scalar{\mathscr{G}^\lambda\eta}{(\mathcal{H}_0+\lambda)\+w_\lambda}[\hilbert_N]\\
    &=\!\scalar{(\mathcal{H}_{\-A}+\lambda)\+\varphi}{\psi}[\hilbert_N]\!-8\pi\+\scalar{v_\lambda|_\pi}{\xi}[\Lp{2}[\pi]]+8\pi\+\scalar{\eta}{w_\lambda|_\pi}[\Lp{2}[\pi]]\\[-2.5pt]
    &=\!\scalar{(\mathcal{H}_{\-A}+\lambda)\+\varphi}{\psi}[\hilbert_N]\!-8\pi\!\textstyle\sum\limits_{i\+<\+j}^N\!\Big[\scalar{(A+\Gamma_{ij}^\lambda)\+\eta}{\xi}[\Lp{2}[\pi_{ij}]]\!-\mspace{-2.25mu}\scalar{\eta}{(A+\Gamma_{ij}^\lambda)\+\xi}[\Lp{2}[\pi_{ij}]]\mspace{-0.75mu}\Big]\\[-5pt]
    &=\!\scalar{(\mathcal{H}_{\-A}+\lambda)\+\varphi}{\psi}[\hilbert_N].
\end{align*}
Furthermore, $\mathcal{H}_{\-A}$ is lower-bounded.
Indeed, let $\psi\in\dom{\mathcal{H}_{\-A}}$
\begin{align*}
\scalar{\psi}{\mathcal{H}_{\-A}\psi}[\hilbert_N]&=\scalar{\psi}{(\mathcal{H}_0+\lambda)\+w_\lambda}[\hilbert_N]-\lambda\norm{\psi}[\hilbert_N]^2\\
&=\scalar{w_\lambda}{(\mathcal{H}_0+\lambda)\+w_\lambda}[\hilbert_N]+\scalar{\mathscr{G}^\lambda\xi}{(\mathcal{H}_0+\lambda)\+w_\lambda}[\hilbert_N]-\lambda\norm{\psi}[\hilbert_N]^2\\
&=\scalar{w_\lambda}{(\mathcal{H}_0+\lambda)\+w_\lambda}[\hilbert_N]+\+\Phi^\lambda[\xi]-\lambda\norm{\psi}[\hilbert_N]^2\!.
\end{align*}
Then
\[
\scalar{\psi}{\mathcal{H}_{\-A}\psi}[\hilbert_N]\geq -\lambda_0\norm{\psi}[\hilbert_N]^2
\]
as soon as $\Phi^\lambda > 0$ for any $\lambda\->\-\lambda_0.$

\n Next, we  prove self-adjointness by showing that $\ran{\mathcal{H}_{\-A}+\lambda}=\hilbert_N$ for $\lambda>\lambda_0.$
Indeed, for any $f\in\hilbert_N$ we 
define $u\in\dom{\mathcal{H}_{\-A}}$  as follows
$$u=w_\lambda+\mathscr{G}^\lambda\xi,\qquad w_\lambda=(\mathcal{H}_0+\lambda)^{-1} f,\qquad \xi = (A+\Gamma_{ij}^\lambda)^{-1}(w_\lambda|_{\pi_{ij}})$$
with $\lambda > \lambda_0$. 
One has
$$(\mathcal{H}_{\-A}+\lambda)\+u=(\mathcal{H}_0+\lambda)\+ w_\lambda=f$$
and therefore $\ran{\mathcal{H}_{\-A}+\lambda}=\hilbert_N$ and self-adjointness is proved.

\n
In conclusion, for any function $A$ (\emph{c.f.}~\eqref{alfa}) we have obtained a s.a. and lower bounded Hamiltonian $\mathcal{H}_{\-A}\mspace{-0.5mu}, \mspace{-0.75mu}\dom{\mathcal{H}_{\-A}\-}$ for the Bose gas, where we recall that the function $A$  depends on the parameters $\alpha \in \R$, $\gamma > \gamma_c$ and on the function $\theta$. 

\n
As a further remark, it is interesting to establish the connection between our Hamiltonian and the one obtained by Albeverio, H{\o}egh-Krohn and Streit in 1977 using the theory of Dirichlet forms~\cite[example 4]{AHK}.
In this work the authors introduce the following Dirichlet form
\begin{equation}\label{AlbeverioDF}
    E[\psi]\vcentcolon=\integrate[\R^{3N}]{\phi^2(\vec{\mathrm{x}})\abs{\nabla \psi(\vec{\mathrm{x}})}^2;\mspace{-10mu}d\vec{\mathrm{x}}},\qquad\psi\in \hilbert_N:\:\nabla \psi \in \Lp{2}[\R^{3N}\-,\,\phi^2(\vec{\mathrm{x}})\+d\vec{\mathrm{x}}],
\end{equation}
where
\begin{equation}\label{weightFunction}
    \phi\mspace{0.75mu}(\vec{x}_1,\ldots,\vec{x}_N)\vcentcolon=\frac{1}{4\pi} \sum_{i\,<\,j}^N\frac{\;e^{-m\,\abs{\vec{x}_i-\+\vec{x}_j}}\!}{\abs{\vec{x}_i\--\vec{x}_j}}\in\Lp*{2}[\R^{3N}],\qquad m\geq 0.
\end{equation}
They prove that the quantity \begin{equation}\label{dirichletQF0}
    \mathcal{Q}_D[\psi]\vcentcolon=E\!\left[\psi/\phi\right]-2m^2\norm{\psi}^2
\end{equation}
defines a singular perturbation of $\mathcal{H}_0\+,\hilbert_N\cap H^2(\R^{3N})$ supported on $\pi$.
More precisely, for any non-negative value of $m$, the quadratic form $\mathcal{Q}_D$ is associated to a bounded from below operator, denoted by $-\Delta_m$ such that
\begin{gather*}
    -\Delta_m \psi=\mathcal{H}_0\+\psi,\qquad \forall \psi\-\in\!\hilbert_N\cap H_0^2(\R^{3N}\setminus \pi),\\
    -\Delta_m\-\geq\- -2\+m^2.
\end{gather*}
In this way the authors define a $N$-body Hamiltonian with contact interactions (according to definition~\ref{def1}) with preassigned lower bound $-2m^2$ and therefore stable.
However, in~\cite{AHK} the domain of the Hamiltonian is not characterized.
In other words, it is not specified which boundary condition on the coincidence hyperplanes is satisfied by an element of the domain of the Hamiltonian and therefore it is not clear which kind of contact interaction is defined.
In order to clarify this point, we have to rewrite $\mathcal{Q}_D$ in our formalism so that an explicit comparison with our results can be made.
In this way we can prove (see~\cite{FeT3}) that the Hamiltonian $-\Delta_m$ defined in~\cite{AHK} is a special case of our family of Hamiltonians $\mathcal{H}_{\-A}$.
More precisely, let $A_0$ be the function~\eqref{alfa} with
\begin{equation*}
    \alpha=-m, \qquad  \gamma=2 \qquad \text{and} \qquad \theta: r\longmapsto e^{-m\+r},\quad r>0.
\end{equation*}
Then one has
\begin{equation*}
    -\Delta_m\-=\mathcal H_{\-A_0} \,.
\end{equation*}
Notice that the Hamiltonian defined in~\cite{AHK} is characterized by a non positive two-body scattering length and by a particular choice of the function $\theta$ explicitly dependent on the two-body scattering length.

\section{Gas with an impurity }

Here we consider the simpler case of a free gas made of $N$ (distinguishable, yet with the same mass) particles interacting with a different test particle via contact interaction.
The Hilbert space of the system is now 
    $L^2(\R^{3(N+1)}) $ and, at a formal level, the Hamiltonian is 
\begin{equation}\label{formalH}
   \mathcal{H}_0 +\sum_{i\+=\+1}^N \nu_i \+\delta(\vec{x}_i-\vec{x}),
\end{equation}

\vspace{-0.3cm}

\n
where 
\begin{equation}\label{freeH}
    \define{\mathcal{H}_0;-\frac{1}{2m\!}\,\Delta_{\vec{x}}- \frac{1}{2 M} \sum_{i\+=\+1}^N \Delta_{\vec{x}_i}  },\qquad\dom{\mathcal{H}_0}= H^2(\R^{3(N+1)}).
\end{equation}
is the free Hamiltonian, $m$ is the mass of the test particle, $M$ is the mass of the particles of the gas and $\nu_i$ are coupling constants.
In order to construct the above Hamiltonian as a s.a. and bounded from below operator in $L^2(\R^{3(N+1)})$ we follow the same line of thought of the case of the interacting Bose gas.
More precisely, we want to construct an operator satisfying condition $(a)$ and the following b.c. at the coincidence hyperplanes 
$$\pi = \textstyle\bigcup\limits_{i=1}^N \pi_i\,, \qquad \pi_i = \{ (\vec{x}, \vec{x}_1,\ldots,\vec{x}_N)\in\R^{3(N+1)} \,\big| \; \vec{x}_i\-=\vec{x} \} $$
\begin{equation}\label{mfBC2}
    \psi(\vec x, \vec X )=\;\frac{\xi_i \!\left(\frac{m\+\vec{x}
    \++M \vec{x}_i}{m\++M},\vec{X}_{\!i}\!\right)\!}{\abs{\vec{x}-\vec{x}_i}}\++(B \+\boldsymbol{\xi})_i \!\left(\tfrac{m\+\vec{x}
    \++M \vec{x}_i}{m\++M},\vec{X}_{\!i}\!\right)\-+\oSmall{1}\-,\quad \abs{\vec{x}\--\vec{x}_i}\to[\!\quad\!]0,
\end{equation}
where we have denoted by $\maps{\xi_i}{\pi_i;\C}$ the charge distributed along $\pi_i$ and by $\boldsymbol{\xi}\-=\-(\xi_1,\ldots,\xi_N)$ their collection, while we have set the shortcuts $\vec{X}\!\-=\-(\vec{x}_1,\ldots,\vec{x}_N)$,  $\vec{X}_{\!i}\!=\!(\vec{x}_1,\ldots,\vec{x}_{i-1},\vec{x}_{i+1},\ldots,\vec{x}_N\-)$ and the operator $B$ in $\bigoplus_{i\+=\+1}^N \-L^2(\pi_i)$  is given by 
\begin{equation}\label{alfaNplus1}
    (B\+\boldsymbol{\xi})_i (\vec{z},\vec{X}_{\!i} ) =  \,\alpha_i\+\xi_i(\vec{z},\vec{X}_{\!i})+\gamma\sum_{\substack{k\+=\+1,\\[1pt] k\+ \neq\+ i}}^{N}\frac{\theta(\abs{\vec{x}_k\--\vec{z}})}{\abs{\vec{x}_k\--\vec{z}}}\,\xi_k(\vec{z},\vec{X}_{\!i})
\end{equation}
where $\alpha_i \in \R$, $\gamma>0$  and the function $\theta$ is essentially bounded and satisfies~\eqref{positiveBoundedCondition}.
We stress that in this case the operator $B$ introduces in~\eqref{alfaNplus1} only an effective repulsive three-body force that makes the contact interaction between $\vec x$ and $\vec x_i$ weaker and weaker as a third particle $\vec x_k$ approaches the common position of $\vec x$ and $\vec x_i$.
As before, one can choose an arbitrarily small, compact support of the function $\theta$ so that the usual two-body b.c. is restored as soon as the other particles of the gas are sufficiently far from $\pi_i$.

\n
As we did for the interacting Bose gas, 
it is convenient to introduce 
the potential
\begin{align}\label{potentialDef2}
    (\mathscr{G}^\lambda\boldsymbol{\xi})(\vec x, \vec X)\vcentcolon&\mspace{-5mu}=\tfrac{2\pi (1+\+\eta)}{m} \sum_{i\+=\+1}^N
    \integrate[\R^{3N}]{\xi_i(\vec{y},\vec{Y}_{\!\!i})\:G^\lambda\-
    \bigg(\!
        \begin{array}{r c l}
            \!\-\vec{x}\+,& \!\!\-\vec{x}_i\+, & \!\!\-\vec{X}_{\!i}\!\\[-2.5pt]
            \!\-\vec{y},& \!\!\-\vec{y},& \!\!\-\vec{Y}_{\!\!i}\!
        \end{array}
        \!\!\bigg); \mspace{-10mu} d\vec{y}d\vec{Y}_{\!\!i}}\\
    &=\vcentcolon\sum_{i\+=\+1}^N \,(\mathscr{G}^\lambda_i\+\xi_i)(\vec x,\vec X),\nonumber
\end{align}
where 
\[
\eta\-=\-\frac{m}{M}>0
\]
is the mass ratio and  
$G^\lambda$ is the kernel of the operator $(\mathcal H_0 +\lambda)^{-1}\!$,  $\lambda\->\-0$, given by 
\begin{equation}\label{greenLaplaceNplus1}
    G^\lambda\-
    \bigg(\!
        \begin{array}{r l}
            \!\-\vec{x}\+,& \!\!\!\-\vec{X}\!\!\\[-2.5pt]
            \!\-\vec{y},& \!\!\!\-\vec{Y}\!\!
        \end{array}
        \!\mspace{-0.75mu}\bigg)\! =\frac{1}{\eta^{\frac{3N}{2}}}\frac{1}{\,(2\pi)^{\frac{3(N+1)}{2}}\!\-}\left(\-\frac{\!\!2m\+\lambda}{\+\abs{\vec{x}\--\-\vec{y}}^2\-+\-\eta\+\abs{\vec{X}\-\!-\-\vec{Y}}^2\!}\-\right)^{\!\frac{3N+1}{4}}\nquad\mathrm{K}_{\frac{3N+1}{2}}\!\-\left(\-\sqrt{2m\lambda}\,\sqrt{\abs{\vec{x}\--\-\vec{y}}^2\-+\-\eta\+\abs{\vec{X}\-\!-\-\vec{Y}}^2}\right)\-.
\end{equation}
Moreover, in the Fourier space we have
\begin{equation}\label{potentialFourierNplus1}
    (\widehat{\mathscr{G}^\lambda_i\+\xi_i })(\vec p, \vec P)= (1+\eta) 
    \sqrt{\frac{2}{\pi} }
    \frac{\FT{\xi}_i(\vec{p}+\vec{p}_i,\vec{P}_{\!i})}{\+p^2 + \eta P^2\-+2m\+\lambda}.
\end{equation}
Finally, the asymptotic behavior near the hyperplane $\pi_i$ is 
\begin{equation} \label{potNplus1}
(\mathscr{G}^\lambda\boldsymbol{\xi})(\vec x, \vec X)=\frac{\xi_i\big(\tfrac{m\+\vec{x}\++M\vec{x}_i}{m\++M}\-,\vec{X}_{\!i}\big)}{\abs{\vec{x}_i\--\vec{x}}}-(\Xi^\lambda \boldsymbol{\xi})_i \big(\tfrac{m\+\vec{x}\++M\vec{x}_i}{m\++M}\-,\vec{X}_{\!i}\big)+\oSmall{1}\-,
\end{equation}
with
\begin{equation}
    \begin{split}
    (\widehat{\Xi^\lambda\boldsymbol{\xi}})_i (\vec{p},\vec{P}_{\!i})=&\,\sqrt{\tfrac{\eta}{(1+\+\eta)^2\!}\,p^2\-+\tfrac{\eta}{1+\+\eta}\+P_{\!i}^2\-+\tfrac{2m}{1+\+\eta}\+\lambda\,}\,\FT{\xi}_i(\vec{p},\vec{P}_{\!i})\,+\\
    &-\frac{1+\eta}{2\pi^2}\sum_{\substack{j\+ = \+ 1,\\[1pt]j\+\neq\+i}}^N\integrate[\R^3]{\frac{\FT{\xi}_j(\vec{p}\--\-\vec{p}_i\-+\-\vec{p}_j,\vec{P}_{\!j})}{\abs{\vec{p}_i-\+\vec{p}}^2\-+ \eta \+P^2\-+2m\lambda}; \!d\vec{p}_i}.
    \end{split}
\end{equation}

\vs

\n
Using the above potential, we decompose an element  $\psi$ of the Hamiltonian domain as
\begin{equation}
    \psi=w_\lambda+\mathscr{G}^\lambda\+\boldsymbol{\xi},\qquad w_\lambda\in H^2(\R^{3(N+1)})
\end{equation}
and consequently the b.c.~\eqref{mfBC} can be rewritten as follows
\begin{equation}
    (B \boldsymbol{\xi} +\Xi^\lambda \boldsymbol{\xi} )_i =w_\lambda|_{\pi_i}.
\end{equation}
The key technical result is  the following.
\begin{prop}\label{crucialResultToProve2}
 Let us fix $\gamma > \hat{\gamma}_c(N,\eta)\,$, where
\begin{equation}\label{criticalGamma}
    \hat{\gamma}_c(N,\eta)= \frac{2(\eta\-+\-1)}{\pi}\arcsin\!\big(\tfrac{1}{\eta\,+\+1}\big)-
    \frac{2  \sqrt{ \eta\+(\eta+2)}}{\pi (N\!-1) (\eta + 1)} \,.
\end{equation}
Then the operator $B+\Xi^\lambda\!,\: \dom{B+\Xi^\lambda}=H^1(\pi)$ acting in $L^2(\pi)$  is s.a. and there exists $\lambda_0>0$ s.t. $B+\Xi^\lambda$ is positive for any $\lambda>\lambda_0.$
\end{prop}

\n
We notice that the critical parameter  $\hat{\gamma}_c$ is positive and
\begin{align*}
    \inf_{\eta\,>\+0}\hat{\gamma}_c(N,\eta)=\frac{2}{\pi}\frac{N\--\-2}{N\--\-1}, && \sup_{\eta\,>\+0}\hat{\gamma}_c(N,\eta)=1.
\end{align*}
Therefore, to simplify the notation from now on we fix $\gamma=1$. 

\n 
Also in this case the result can be proved by considering the quadratic form in $\Lp{2}[\pi]$
\begin{equation}
\Phi^\lambda[\boldsymbol{\xi}]=\textstyle\sum\limits_{i\+=\+1}^N\scalar{\xi_i}{ 
(B \boldsymbol{\xi} +\Xi^\lambda \boldsymbol{\xi} )_i } [\Lp{2}[\pi_i]],\qquad\dom{\Phi^\lambda}=H^{\frac{1}{2}}(\pi)
\end{equation}
and by showing that $\Phi^\lambda$ is closed for any $\lambda\->\-0$ and bounded from below by a positive constant when $\lambda\->\-\lambda_0$ (for the proof see~\cite{FeT}).  
Once the proposition~\ref{crucialResultToProve2} is proven, one defines the operator 
\begin{defn}\label{Nplus1Hamiltonian}
    \begin{equation*}
        \begin{dcases}
            \dom{\mathcal{H}_{\-B}}\-=\mspace{-2.25mu}\Big\{\psi\-\in\mspace{-2.25mu}\Lp{2}[\R^{3(N+1)}]\:\big|\;\psi\mspace{-0.75mu}-\mathscr{G}^\lambda\boldsymbol{\xi}\-=\-w_\lambda\!\in \mspace{-2.25mu}H^2(\R^{3(N+1)}),\, \boldsymbol{\xi}\-\in\mspace{-2.25mu} H^1(\pi),\:(B \boldsymbol{\xi}\-+\- \Xi^\lambda \boldsymbol{\xi})_i\- =\mspace{-0.75mu}w^\lambda|_{\pi_i},\,\lambda\->\-0\Big\}\\
            \mathcal{H}_{\-B}\+\psi=\mathcal{H}_0 \+ w_\lambda-\lambda\+\mathscr{G}^\lambda\boldsymbol{\xi}.
        \end{dcases}
    \end{equation*}
\end{defn}
\n
Following the line of previous case, it is now straightforward to show that such operator is s.a. and bounded from below in $L^2(\R^{3(N+1)})$. We can also write an explicit representation for the resolvent. For any $f \!\in\! L^2 (\R^{3(N+1)} )$ and $\lambda\- >\-\lambda_0$ we have
\begin{equation}\label{resol1}
( \mathcal{H}_{\-B} + \lambda )^{-1} f=  (\mathcal{H}_0+\lambda)^{-1}f+\mathscr{G}^\lambda\boldsymbol{\xi}
\end{equation}
where $\boldsymbol{\xi}$ is the solution of the equation
\begin{equation}\label{resol2} 
(B \boldsymbol{\xi} + \Xi^{\lambda} \boldsymbol{\xi})_i = [(\mathcal{H}_0+\lambda)^{-1}f]\big|_{\pi_i} . 
\end{equation}
To be more specific, equation~\eqref{resol2} is explicitly written as follows 
\begin{equation}\begin{split}\label{resol22}
\sum_{j\+=\+1}^N  \!&\left(\- \alpha_i\+ \delta_{ij} + \frac{\theta(|\vec x_i - \vec x_j|)}{|\vec x_i - \vec x_j |} \+(1- \delta_{ij})\! \right) \xi_j (\vec x_j, \vec X_{\!j})+ \\
&+\- \tfrac{1}{(2 \pi\-)^{\frac{3N}{2}} }  \!
\integrate[\R^{3N}]{e^{i \+ \vec x_i \cdot\+ \vec p\, +\,i\+ \vec X_{\!i}\+ \cdot\+ \vec P_{\!i}} ;\mspace{-10mu}d\vec p \+d\mspace{-0.75mu} \vec P_{\!i}}
\Bigg[  \sqrt{ \tfrac{\eta}{(1 + \eta)^2} p^2 + \tfrac{\eta}{1 + \eta} P_{\!i}^2 + \tfrac{2m \lambda}{1 + \eta} \,}\,  \hat{\xi}_i ( \vec p, \vec P_{\!i})\++ \\ 
&- \tfrac{1+\eta}{2 \pi^2} \sum_{\substack{j\+=\+1,\\[1pt]  j\+\neq\+ i} }^N \integrate[\R^3]{\frac{ \hat{\xi}_j ( \vec p - \vec p_i \-+ \vec p_j, \vec P_{\!j})}{|\vec p_i\- - \vec p|^2 + \eta P^2  + 2 m \lambda};\! d \vec p_i} \Bigg]\! = \left[ (\mathcal H_0 + \lambda)^{-1} f \right]\! (\vec x_i, \vec X)\,.
\end{split}
\end{equation}

\section{The infinite mass limit}

Here we discuss the limit of the model defined in the previous section when the  particles of the gas are infinitely heavy, \emph{i.e.}, when $\eta \to 0$. 
We expect that the kinetic energy of the heavy particles is negligible in the limit and then the coordinates of the heavy particles become fixed parameters.
As a result, one obtains the Hamiltonian for the light particle subject to $N$ fixed point interactions located at $\{\vec{x}_i\}_{i\+=\+1}^N$, namely
\begin{equation}\label{NfixedPointInteractionsHamiltonian}
        \begin{dcases}
            & \nquad \dom{h_{\underline{\alpha}, \theta}}\!= \mspace{-2.25mu}\Big\{\psi\-\in\mspace{-2.25mu}\Lp{2}[\R^{3}]\:\big|\;\psi-\! {\textstyle\sum\limits_{i\+=\+1}^N} g^\lambda(\+ \cdot - \vec x_i) \+ q_i \- =\-w_\lambda\!\in \mspace{-2.25mu}H^2(\R^{3}),\, q_i \-\in \C ,\\[-5pt]
            & \mspace{171mu}\big(\alpha_i \-+\- \sqrt{\mspace{-0.75mu}\lambda}\mspace{0.75mu}\big)\+  q_i  \-+\!\! {\textstyle\sum\limits_{\substack{j\+=\+1,\\ j \+\neq\+ i}}^N} \-\delta_{\lambda,\+\theta}(\vec{x}_i\--\vec{x}_j)\+ q_j \-= w_{\lambda} (\vec x_i) 
            \-\Big\}\!\\[-15pt]
            & \nquad h_{\underline{\alpha},\+ \theta} \+ \psi=h_0 \+ w_\lambda- \lambda\+ {\textstyle\sum\limits_{i\+=\+1}^N} \,g^\lambda (\+\cdot - \vec x_i) \+ q_i \+,
        \end{dcases}
    \end{equation}
    where $\underline{\alpha}=(\alpha_1,\ldots,\alpha_N\-)$ and we have defined the functions
    \begin{align*}
        \delta_{\lambda,\+\theta}:\:\vec{r}\,\longmapsto\+\frac{\theta(\vec{r}) - e^{-\sqrt{\lambda}\, \abs{\vec{r}}}}{\abs{\vec{r}}}, && g^\lambda\-:\: \vec r\,\longmapsto \+\frac{e^{- \sqrt{\lambda}\, \abs{\vec r}}}{\abs{\vec r}}.
    \end{align*}
    \n
    Notice that the b.c. contained  in $\dom{h_{\underline{\alpha}, \theta}}$ is equivalent to the following behavior for $\abs{\vec{x} - \vec{x}_i} \to 0$
    \begin{equation}
    \psi(\vec x) = \frac{q_i}{\abs{\vec{x} \--\- \vec{x}_i}} + \alpha_i\mspace{2.25mu} q_i + \sum_{\substack{j\+=\+1, \\[1pt] j\+\neq\+ i}}^N \frac{\theta(\vec{x}_i\!-\-\vec{x}_j)}{\abs{\vec{x}_i\!-\-\vec{x}_j}} \, q_j + \oSmall{1} \-.
    \end{equation}

\begin{prop}
    Let $\+\mathcal{H}_{\-B}\+$ and $\+h_{\underline{\alpha},\+ \theta}\+$ be given by definition~\ref{Nplus1Hamiltonian} and equation~\eqref{NfixedPointInteractionsHamiltonian}, respectively.
    Then, one has
    \begin{equation}
    \lim_{\eta\to 0}\mathcal{H}_{\-B}= h_{\underline{\alpha},\+ \theta}\otimes \Id_{\Lp{2}[\R^{3N}\!,\, d\vec{x}_1\mspace{-0.75mu}\cdots \+d\vec{x}_{\-N}\-]}
    \end{equation}
    in the strong-resolvent sense.
    \begin{proof}
        Let us fix for simplicity $m=1/2$ and let us first show that one has
        \begin{align}
        \lim_{\eta\to 0}\left[ (\mathcal{H}_0\-+\-\lambda)^{-1}\-f \right]\! (\vec{x}, \mspace{-0.75mu}\vec{X}\mspace{-0.75mu}) &= \frac{1}{4\pi}\!\-\integrate[\R^3]{g^\lambda(\vec{x} \--\- \vec{y}) f(\vec y, \mspace{-0.75mu}\vec{X}\mspace{-0.75mu}); \!d\vec y}\qquad\text{in }\,\Lp{2}[\R^{3(N+1)}]\\
        &= [(h_0\- +\- \lambda)^{-1}\!\otimes\-\Id_{\Lp{2}[\R^{3N}]}\+ f] (\vec{x}, \mspace{-0.75mu}\vec{X}\mspace{-0.75mu}).\nonumber
        \end{align}
        Indeed, by dominated convergence theorem one proves that for any $f\!\in\!\Lp{2}[\R^{3(N+1)}]$
        $$\frac{\FT{f}(\vec{p}, \vec{P})}{p^2+\eta P^2+\lambda}\xlongrightarrow[\eta\to 0]{} \frac{\FT{f}(\vec{p}, \vec{P})}{p^2+\lambda}\qquad\text{in }\,\Lp{2}[\R^{3(N+1)}\!,\,d\vec{p}\+d\mspace{-0.75mu}\vec{P}]\,.$$
        Moreover
        \begin{align*}
            \integrate[\R^{3(N\-+1)}]{\frac{e^{i\+\vec{p}\+\cdot\+\vec{x}\,+\,i\mspace{0.75mu}\vec{P}\mspace{0.75mu}\cdot\vec{X}}\!\-}{(2\pi)^{\frac{3(N\-+1)}{2}}}\,\frac{\FT{f}(\vec{p},\vec{P})}{p^2+\lambda};\mspace{-36mu}d\vec{p}\+d\mspace{-0.75mu}\vec{P}}&=
            \frac{1}{(2\pi)^3}\!\-\integrate[\R^3]{f(\vec{y},\vec{X});\!d\vec{y}}\!\-\integrate[\R^3]{\frac{e^{-i\+\vec{p}\+\cdot(\vec{y}-\vec{x})}}{p^2+\lambda};\!d\vec{p}}\\
            &=\frac{1}{4\pi}\!\integrate[\R^3]{\frac{e^{-\sqrt{\lambda}\,\abs{\vec{x}-\vec{y}}}}{\abs{\vec{x}\--\-\vec{y}}}\,f(\vec{y},\vec{X});\!d\vec{y}}.
        \end{align*}
        Similarly, from~\eqref{potentialFourierNplus1} we also have
        \begin{equation}
        (\mathcal G^{\lambda} \boldsymbol{\xi}) (\vec x, \vec X)= \sum_{i\+=\+1}^N (\mathcal G^{\lambda}_i \xi_i) (\vec x, \vec X) \xlongrightarrow[\eta\to 0]{} \sum_{i\+=\+1}^N g^\lambda (\vec x- \vec x_i)  \, \xi_i(\vec{x}_i,\vec{X}_{\!i})\qquad\text{in }\,\Lp{2}[\R^{3(N+1)}],
        \end{equation}
        since one can find a uniform majorant which is integrable in $\R^{3(N+1)}$, \emph{i.e.}
        $$\frac{2}{\pi}\+\abs{\FT{\xi}_i(\vec{p}+\vec{p}_i,\vec{P}_{\!i})}^2\-\left[\frac{1}{p^2\-+\-\eta P^2\-+\-\lambda}-\frac{1}{p^2\-+\-\lambda}\right]^2\!\!\leq \frac{2}{\pi}\frac{\abs{\FT{\xi}_i(\vec{p}+\vec{p}_i,\vec{P}_{\!i})}^2\!\-}{(p^2\-+\-\lambda)^2}.$$
        Hence the limit can be straightforwardly computed
        \begin{align*}
        \sqrt{\frac{2}{\pi}}\!\integrate[\R^{3(N\-+1)}]{\frac{e^{i\+\vec{p}\+\cdot\+\vec{x}\,+\,i\mspace{0.75mu}\vec{P}\mspace{0.75mu}\cdot\vec{X}}\!\-}{(2\pi)^{\frac{3(N\-+1)}{2}}}\,\frac{\FT{\xi}_i(\vec{p}+\vec{p}_i,\vec{P}_{\!i})}{p^2+\lambda};\mspace{-36mu}d\vec{p}\+d\mspace{-0.75mu}\vec{P}}&=\sqrt{\frac{2}{\pi}}\!\integrate[\R^{3(N\-+1)}]{\frac{e^{i\+\vec{p}\+\cdot(\vec{x}-\vec{x}_i)\,+\,i\+\vec{q}\+\cdot\+\vec{x}_i\,+\,i\mspace{0.75mu}\vec{P}_{\!i}\mspace{0.75mu}\cdot\vec{X}_{\!i}}\!\-}{(2\pi)^{\frac{3(N\-+1)}{2}}}\,\frac{\FT{\xi}_i(\vec{q},\vec{P}_{\!i})}{p^2+\lambda};\mspace{-36mu}d\vec{p}\+d\vec{q}d\mspace{-0.75mu}\vec{P}_{\!i}}\\
        &=\sqrt{\frac{2}{\pi}}\,\xi_i(\vec{x}_i,\vec{X}_{\!i})\!\-\integrate[\R^3]{\frac{e^{i\+\vec{p}\+\cdot(\vec{x}-\vec{x}_i)}\!\-}{(2\pi)^{3/2}}\,\frac{1}{p^2+\lambda};\!d\vec{p}}\\
        &=g^\lambda(\vec{x}\--\-\vec{x}_i)\+\xi_i(\vec{x}_i,\vec{X}_{\!i}).
        \end{align*}
        Moreover, from~\eqref{resol22} one sees that the charges $\xi_i(\vec x_i, \vec X_{\!i})$ reduce for $\eta \to 0$ to the (constant with respect to $\vec{x}$) charges $q_i= q_i(\vec x_i, \vec X_{\!i})$ solving the linear system 
        \begin{equation}\label{eqq}
         \big(\alpha_i \-+\- \sqrt{\lambda} \+\big)\+  q_i  +\! \sum_{\substack{j\+=\+1,\\ j\+ \neq\+ i}}^N  \delta_{\lambda,\+\theta}(\vec{x}_i\--\-\vec{x}_j) \+  q_j = [(h_0+\lambda)^{-1}\-f] (\vec x_i).
        \end{equation}
        Here we are adopting a slight abuse of notation by dropping the parametric dependence on $\vec{X}$.
        In order to prove~\eqref{eqq} let us compute the limit for $\eta\to 0$ of the l.h.s. of~\eqref{resol22}.
        First of all, exploiting the elementary inequality $\sqrt{a+b}-\sqrt{b}\leq\sqrt{a}$
        \begin{align*}
        \integrate[\R^{3N}]{\Big(\sqrt{\tfrac{\eta}{(1+\eta)^2}p^2\-+\-\tfrac{\eta}{1+\eta}P_{\!i}^2\-+\-\tfrac{\lambda}{1+\eta}}-\sqrt{\lambda}\+\Big)^{\!2}\+\abs{\FT{\xi}_i(\vec{p},\vec{P}_{\!i})}^2;\mspace{-10mu}d\vec{p}\+d\mspace{-0.75mu}\vec{P}_{\!i}}&\leq\!\-\integrate[\R^{3N}]{(\eta\+ p^2\-+\-\eta P_{\!i}^2)\+\abs{\FT{\xi}_i(\vec{p},\vec{P}_{\!i})}^2;\mspace{-10mu}d\vec{p}\+d\mspace{-0.75mu}\vec{P}_{\!i}}\\
        &\leq \eta\norm{\xi_i}[H^1(\R^{3N})]^2\-,
        \end{align*}
        hence
        \begin{equation}\label{diagonalEtaSmall}
            \lim_{\eta\to 0}\sqrt{\tfrac{\eta}{(1+\eta)^2}p^2\-+\-\tfrac{\eta}{1+\eta}P_{\!i}^2\-+\-\tfrac{\lambda}{1+\eta}\+}\+\FT{\xi}_i(\vec{p},\vec{P}_{\!i})=\sqrt{\lambda\+}\+\FT{\xi}_i(\vec{p},\vec{P}_{\!i}),\quad\text{in }\,\Lp{2}[\R^{3N}].
        \end{equation}
        Concerning the remaining term one can verify the convergence by finding a suitable uniform majorant in $\R^{3(N+2)}$.
        To this end, set
        $$\FT{f}_{ij}(\vec{q},\vec{k},\vec{P}_{\!ij}\mspace{-0.75mu})\vcentcolon=\FT{\xi}_j(\vec{q},\vec{p}_1,\ldots,\underset{i\text{-th}}{\vec{k}},\ldots,\vec{p}_{j-1},\vec{p}_{j+1},\ldots,\vec{p}_N\-),$$
        so that
        \begin{align*}
            \integrate[\R^{3N}]{;\mspace{-10mu}d\vec{p}\+d\mspace{-0.75mu}\vec{P}_{\!i}}\!\left\lvert\integrate[\R^3]{\FT{\xi}_j(\vec{p}\--\-\vec{p}_i\!+\-\vec{p}_j,\vec{P}_{\!j})\bigg(\frac{1}{\abs{\vec{p}\--\-\vec{p}_i}^2\!+\-\eta P^2\!+\-\lambda}-\frac{1}{\abs{\vec{p}\--\-\vec{p}_i}^2\!+\-\lambda}\bigg)\-;\!d\vec{p}_i}\right\rvert^2\\
            \leq \!\-\integrate[\R^{3(N\-+2)}]{\frac{\abs{\FT{f}_{ij}(\vec{p}\--\-\vec{k}\!+\-\vec{p}_j,\vec{k},\vec{P}_{\!ij})}\+\abs{\FT{f}_{ij}(\vec{p}\--\-\vec{k}'\!\-+\-\vec{p}_j,\vec{k}'\!,\vec{P}_{\!ij})}}{(\abs{\vec{p}\--\-\vec{k}}^2\!+\-\lambda)(\abs{\vec{p}\--\-\vec{k}'}^2\!+\-\lambda)};\mspace{-36mu}d\vec{p}\+d\vec{p}_j d\mspace{-0.75mu}\vec{P}_{\!ij}d\vec{k}\+d\vec{k}'}\\
            \leq \!\-\integrate[\R^9]{\frac{\FT{h}_{ij}(\vec{k})\FT{h}_{ij}(\vec{k}')}{\abs{\vec{p}\--\-\vec{k}}^2\+\abs{\vec{p}\--\-\vec{k}'}^2};\!d\vec{p}\+d\vec{k}\+d\vec{k}'}
        \end{align*}
        where we have used a Cauchy-Schwarz inequality and we have set $\displaystyle \FT{h}_{ij}(\vec{k})\-\vcentcolon=\-\sqrt{\-\integrate[\R^{3(N-1)}]{\abs{\FT{f}_{ij}(\vec{q},\vec{k},\vec{P}_{\!ij}\-)}^2;\mspace{-36mu}d\vec{q}\+d\mspace{-0.75mu}\vec{P}_{\!ij}}}.$
        We need to show that the r.h.s. of the latter inequality is finite.
        %In particular, let us adopt the following Cauchy-Schwarz inequality
        %$$\leq \!\-\integrate[\R^9]{\frac{\FT{h}_{ij}(\vec{k})\FT{h}_{ij}(\vec{k}')}{\abs{\vec{p}\--\-\vec{k}}^2\+\abs{\vec{p}\--\-\vec{k}'}^2};\!d\vec{p}\+d\vec{k}\+d\vec{k}'},$$
        %where $\displaystyle \FT{h}_{ij}(\vec{k})\vcentcolon=\sqrt{\integrate[\R^{3(N-1)}]{\abs{\FT{f}_{ij}(\vec{q},\vec{k},\vec{P}_{\!ij}\-)}^2;\mspace{-36mu}d\vec{q}\+d\mspace{-0.75mu}\vec{P}_{\!ij}}}.$
        Integrating along $\vec{p}$ one obtains
        \begin{align*}
            \integrate[\R^3]{\frac{1}{\abs{\vec{p}\--\-\vec{k}}^2\+\abs{\vec{p}\--\-\vec{k}'}^2};\!d\vec{p}}&=2\pi\!\-\integrate[0;\pInfty]{;\nquad dp}\!\-\integrate[-1;1]{\frac{1}{p^2\-+\-\abs{\vec{k}\--\-\vec{k}'}^2\-+\-2p\+\abs{\vec{k}\--\-\vec{k}'}\+u};du}\\
            &=\frac{2\pi}{\abs{\vec{k}\--\-\vec{k}'}}\-\integrate[0;\pInfty]{\frac{1}{p}\ln\!\left(\frac{p+\abs{\vec{k}\--\-\vec{k}'}}{\abs{p-\abs{\vec{k}\--\-\vec{k}'}}}\right);\nquad dp}\!=\frac{\pi^3}{\abs{\vec{k}\--\-\vec{k}'}}.
        \end{align*}
        Therefore, according to Hardy's inequality
        \begin{align*}
            \integrate[\R^9]{\frac{\FT{h}_{ij}(\vec{k})\FT{h}_{ij}(\vec{k}')}{\abs{\vec{p}\--\-\vec{k}}^2\+\abs{\vec{p}\--\-\vec{k}'}^2};\!d\vec{p}\+d\vec{k}\+d\vec{k}'\-}\-=\pi^3\!\!\integrate[\R^6]{\frac{\FT{h}_{ij}(\vec{k})\FT{h}_{ij}(\vec{k}')}{\abs{\vec{k}\--\-\vec{k}'}};\!d\vec{k}\+d\vec{k}'\-}\-=4\pi^4\!\!\integrate[\R^3]{\frac{\abs{h_{ij}(\vec{x})}^2\!\-}{\abs{\vec{x}}^2};\!d\vec{x}}\\
            \leq 16\pi^4\norm{h_{ij}}[H^1(\R^3)]^2\leq 16\pi^4\norm{\xi_j}[H^1(\R^{3(N+1)})]^2\-.
        \end{align*}
        Thus, the convergence in $\Lp{2}[\R^{3N}]$ has been proven by dominated convergence theorem.
        Additionally, one has
        \begin{align*}
            -\frac{1}{\+2\pi^2\!\-}\-\integrate[\R^{3N}]{\frac{e^{i\+\vec{p}\+\cdot\+\vec{x}_i\,+\,i\+\vec{P}_{\!i}\mspace{0.75mu}\cdot\vec{X}_{\!i}}\!\-}{(2\pi)^{\frac{3N}{2}}};\mspace{-10mu}d\vec{p}\+d\mspace{-0.75mu}\vec{P}_{\!i}}\!\!\integrate[\R^3]{\frac{\FT{\xi}_j(\vec{p}\--\-\vec{p}_i\-+\-\vec{p}_j,\vec{P}_{\!j})}{\abs{\vec{p}\--\-\vec{p}_i}^2+\lambda};\!d\vec{p}_i}=\\
            = -\frac{1}{\+2\pi^2\!\-}\-\integrate[\R^{3N}]{\frac{e^{i\+\vec{q}\+\cdot\+\vec{x}_j\,+\,i\+\vec{P}_{\!j}\mspace{0.75mu}\cdot\vec{X}_{\!j}}\!\-}{(2\pi)^{\frac{3N}{2}}}\,\FT{\xi}_j(\vec{q},\vec{P}_{\!j});\mspace{-10mu}d\vec{q}\+d\mspace{-0.75mu}\vec{P}_{\!j}}\!\!\integrate[\R^3]{\frac{e^{i\+\vec{k}\+\cdot(\vec{x}_i-\+\vec{x}_j)}\!\-}{k^2+\lambda};\!d\vec{k}}\\
            =-g^\lambda(\vec{x}_i\!-\mspace{-0.75mu}\vec{x}_j)\+\xi_j(\vec{x}_j,\vec{X}_{\!j}).
        \end{align*}
        In other words, we have established the convergence in $\Lp{2}[\R^{3N}]$ of the l.h.s. of equation~\eqref{resol22}.
        Owing to the convergence in $\Lp{2}[\R^{3(N+1)}]$ of the free resolvent, the continuity of the trace operator and the uniqueness of the limit, we have proven~\eqref{eqq}.
        In conclusion, for $\eta \to 0$ the resolvent~\eqref{resol1} reduces to the resolvent in $L^2(\R^3)$ (\emph{i.e.}, dropping the parametric dependence on $\vec X$) of the Schr\"odinger operator $h_{\underline{\alpha},\+ \theta}$ with {\em non-local} point interactions placed at $\vec x_1, \ldots, \vec x_N$ given by 
        \begin{equation} \label{resol3}
            \left[ (h_{\underline{\alpha},\+ \theta} + \lambda)^{-1}\- f \right]\! (\vec x) = [(h_0+\lambda)^{-1}\- f] (\vec x)  + \sum_{i\+=\+1}^N g^\lambda (\vec x - \vec x_i)  \, q_i    
        \end{equation}
        where $\underline{\alpha} = (\alpha_1,\ldots , \alpha_N)$ and the charges $q_i$ are solutions of equation~\eqref{eqq} (see~\cite{FST}).

    \end{proof}
\end{prop}

\n
The property of non-locality of the point interactions we have obtained is due to the presence of $\theta(\vec{x}_i\!-\-\vec{x}_j)$.
Choosing $\theta$ with small compact support, the property of locality is restored as soon as the distance between the points $\vec x_1, \ldots , \vec x_N$ is sufficiently large.
On the other hand we stress that the presence of $\theta(\vec{x}_i\!-\-\vec{x}_j)$ plays an important role.
In particular it makes the behavior of the model Hamiltonian regular and physically reasonable when two or more of the points $\vec x_1, \ldots , \vec x_N$ approach each other.
To verify this fact, we consider for simplicity the case $N\!=\-2$.
Let us denote $R=\abs{\vec{x}_1 \!-\- \vec{x}_2}$ and let $h_{\underline{\alpha}, \+\theta}(R)$ be the corresponding Hamiltonian.
Then we consider the resolvent~\eqref{resol3} for $R \to 0$ with the point $\vec x_1$ fixed.
The result is the following 
\begin{prop}
Assume that $\theta$ is differentiable in zero with $\theta'(0)=0$ and $\alpha_1 + \alpha_2 \neq 0$.
Then
\begin{equation}
\lim_{R \to 0} h_{\underline{\alpha},\+ \theta} (R) = \mathfrak{h}_{\alpha,\+\vec{x}_1}
\end{equation}
in the norm-resolvent sense, where $\mathfrak{h}_{\alpha,\+\vec{x}_1}$ is the Hamiltonian in $\Lp{2}[\R^3]$ with a point interaction in the fixed point $\vec{x}_1$ with strength $\alpha$
given by
\begin{equation}
    \frac{1}{\alpha}= \frac{1}{\alpha_1} + \frac{1}{\alpha_2} \,.
\end{equation}
    \begin{proof}
    For any $f \in L^2(\R^3)$ we have
    \begin{equation}\label{difres}\begin{split}
        (h_{\underline{\alpha},\+\theta} (R)\- + \lambda)^{-1}\- f \!-\- (\mathfrak{h}_{\alpha,\+\vec{x}_1}\! +\lambda)^{-1}\- f \-=& \+\Big( q_1\- + q_2\- - \frac{[(h_0+\lambda)^{-1}\-f] (\vec x_1) }{\alpha + \sqrt{\lambda}} \Big)\+ g^\lambda (\+\cdot - \vec x_1)\,+ \\
        & + q_2 \Big( g^\lambda (\+\cdot - \vec x_2) - g^\lambda (\+\cdot  - \vec x_1)\- \Big) \+.
    \end{split}
    \end{equation}
    The explicit expression of $q_1$ and $q_2$ can be computed by solving equation~\eqref{eqq} for $N\-=2$.
    One finds
    \begin{gather}
    q_1= \frac{ \big(\alpha_2 \-+\- \sqrt{\mspace{-0.75mu}\lambda} \mspace{0.75mu}\big) [(h_0\mspace{-2.25mu}+\-\lambda)^{-1}\-f] (\vec x_1) \-- \delta_{\lambda,\+\theta}(R)\+ [(h_0\mspace{-2.25mu}+\-\lambda)^{-1}\-f] (\vec x_2) }{ \big(\alpha_1\- +\- \sqrt{\mspace{-0.75mu}\lambda} \mspace{0.75mu}\big)\-\big(\alpha_2\- +\- \sqrt{\mspace{-0.75mu}\lambda} \mspace{0.75mu} \big) \-- \delta_{\lambda,\+\theta}(R)^2 },\\
    q_2= \frac{ \big(\alpha_1 \-+\- \sqrt{\mspace{-0.75mu}\lambda} \mspace{0.75mu}\big) [(h_0\mspace{-2.25mu}+\-\lambda)^{-1}\-f] (\vec x_2) \-- \delta_{\lambda,\+\theta}(R)\+ [(h_0\mspace{-2.25mu}+\-\lambda)^{-1}\-f] (\vec x_1) }{ \big(\alpha_1\- +\- \sqrt{\mspace{-0.75mu}\lambda} \mspace{0.75mu}\big)\-\big(\alpha_2\- +\- \sqrt{\mspace{-0.75mu}\lambda} \mspace{0.75mu} \big) \-- \delta_{\lambda,\+\theta}(R)^2 },\label{secondCharge}\\
    \delta_{\lambda,\+\theta}(R)= \frac{ \theta(R) - e^{-\sqrt{\lambda}R}}{R} =  \sqrt{\lambda} + \oBig{R},\qquad R\to 0\nonumber
    \end{gather}
    and therefore
    \begin{equation*}\begin{split}
    q_1\- + q_2 =&\, \frac{ \big(\alpha_2 \!+\!\sqrt{\mspace{-0.75mu}\lambda} - \delta_{\lambda,\+\theta}(R)\- \big)}{ \big(\alpha_1\- + \-\sqrt{\mspace{-0.75mu}\lambda}\mspace{0.75mu} \big) \big( \alpha_2 + \sqrt{\mspace{-0.75mu}\lambda}\mspace{0.75mu}\big) - \delta_{\lambda,\+\theta}(R)^2}\+[(h_0\-+\-\lambda)^{-1}\- f]  (\vec{x}_1\mspace{-0.75mu})\,+\\
    &+\frac{ \big(\alpha_1 \!+\!\sqrt{\mspace{-0.75mu}\lambda} - \delta_{\lambda,\+\theta}(R) \-\big) }{ \big(\alpha_1\- + \-\sqrt{\mspace{-0.75mu}\lambda}\mspace{0.75mu} \big) \big( \alpha_2 + \sqrt{\mspace{-0.75mu}\lambda}\mspace{0.75mu}\big) - \delta_{\lambda,\+\theta}(R)^2}\+[(h_0\-+\-\lambda)^{-1}\- f]  (\vec{x}_2\mspace{-0.75mu}).\\
    \end{split}\end{equation*}
    For the first term in the r.h.s. of~\eqref{difres} we have
    \begin{align*}
    q_1\- + q_2 - \frac{[(h_0\-+\-\lambda)^{-1}\-f] (\vec{x}_1\mspace{-0.75mu}) }{\alpha + \sqrt{\lambda}}&= \frac{  (1+\oBig{R})[(h_0\-+\-\lambda)^{-1}\- f] (\vec{x}_1\mspace{-0.75mu})}{ \alpha\-+\-\sqrt{\lambda} + \oBig{R} }+
    \\
    & \mspace{-7mu}+ \frac{ ( \tfrac{\alpha_1}{\alpha_1 +\mspace{2.25mu} \alpha_2 }\mspace{-2.25mu} +\mspace{-0.75mu} \oBig{R} \-) \big( [(h_0\mspace{-2.25mu}+\!\lambda)^{-1}\mspace{-2.25mu}f](\vec{x}_2\mspace{-0.75mu})\mspace{-2.25mu} -\- [(h_0\mspace{-2.25mu}+\!\lambda)^{-1}\mspace{-2.25mu}f] (\vec{x}_1\mspace{-0.75mu}) \-\big)\! }{ \alpha\-+\-\sqrt{\lambda} + \oBig{R} }\mspace{-0.75mu}- \frac{ [(h_0\mspace{-2.25mu}+\!\lambda)^{-1}\mspace{-2.25mu}f](\vec{x}_1\mspace{-0.75mu}) }{ \alpha +  \sqrt{\lambda} } \\
    &=
    \oBig{R} \norm{f}[\Lp{2}[\R^3]] + \frac{\frac{\alpha_1}{\alpha_1+\mspace{2.25mu}\alpha_2}}{\alpha\-+\-\sqrt{\mspace{-0.75mu}\lambda}\,}\, \big( [(h_0\-+\-\lambda)^{-1}\-f](\vec{x}_2\mspace{-0.75mu}) - [(h_0\-+\-\lambda)^{-1}\-f] (\vec{x}_1\mspace{-0.75mu}) \big).
    \end{align*}
    It is now sufficient  to recall that because of the boundedness of the resolvent and the Sobolev embedding $H^2(\R^3)\hookrightarrow C^{\+0,\frac{1}{2}\-}(\R^3)$ 
    one has for some $c>0$
    $$\left\lvert[(h_0\-+\-\lambda)^{-1}\- f](\vec x)- [(h_0\-+\-\lambda)^{-1} \-f](\vec x')\right\rvert < c \, \abs{\vec x \--\- \vec x'}^{\frac{1}{2}} \norm{f}[\Lp{2}[\R^3]]\-,$$
    thus we can conclude that the first term in the r.h.s. of~\eqref{difres} goes to zero for $R \to 0$ uniformly in $f\-\in\-\Lp{2}[\R^3]$. 
    
    \n
    Concerning the second term in the r.h.s. of~\eqref{difres}, by the means of~\eqref{secondCharge} we first notice that $\abs{q_2} < c \+ \norm{f}[\Lp{2}[\R^3]]$. Moreover, we have
    \begin{align*}
    \norm{g^{\lambda} (\cdot - \vec x_2) - g^{\lambda} (\cdot  - \vec x_1)}[\Lp{2}[\R^3]]^2 \! &= \frac{2}{\pi}\- \integrate[\R^3]{\!\left\lvert \frac{ e^{i\+ \vec p \+\cdot\+ \vec x_2} - e^{i\+ \vec p\+ \cdot\+ \vec x_1} }{p^2 \-+\- \lambda} \right\rvert^2;\!d\vec{p}} \\
    &= \frac{2}{\pi}\- \integrate[0;\pInfty]{\frac{p^2}{(p^2 + \lambda)^2}; \nquad dp}\!\! \integrate[-1;1]{\!\left\lvert 1\- - e^{i\+ p R \+u} \right\rvert^2;du}\\
    &= \frac{8}{\pi}\- \integrate[0;\pInfty]{\frac{p^2}{(p^2 + \lambda)^2}\left( 1 - \frac{ \sin pR}{pR} \right); \nquad dp}
    \end{align*}
    which goes to zero for $R \to 0$ by dominated convergence theorem. This concludes the proof.

\end{proof}
\end{prop}

\n
Some comments are in order. The proposition shows that in the limit $R \to 0$ one obtains a point interaction with a scattering length $\alpha^{-1}$ equal to the sum of the scattering lengths of the two original point interactions (considered separately from each other). In particular, if one or both of the $\alpha_i$ are zero then also $\alpha =0$, which means infinite scattering length. 
We stress that this is the right behavior of the model one expects from the physical point of view (see also~\cite{FST}). 

\n
Notice that if $\alpha_1\- + \alpha_2=0$ then it is immediate to see that in the limit one  obtains the free Hamiltonian. 
Finally, if we assume that $\theta'(0) \neq 0$ then in the limit we find a point interaction with
\begin{equation}\label{alli}
\alpha = \frac{ \alpha_1 \alpha_2 - \theta'(0)^2}{ \alpha_1 + \alpha_2 - 2\+ \theta'(0)}\,. 
\end{equation}

\n
It is  worth mentioning that the limit  behavior for $R \to 0$ of our {\em non-local} point interactions  is completely different for the standard {\em local} point interactions, \emph{i.e.} when $\theta$ is identically zero. Indeed, in such a case one has 
\begin{equation}
\delta_{\lambda,\+\theta}(R) = -\frac{1}{R} + \sqrt{\lambda} + \oBig{R}\-,\qquad R\to 0
\end{equation}
and it is easy to check that the resolvent~\eqref{difres} reduces to the free resolvent for $R \to 0$.
In other words, the point interactions simply disappear in the limit and this is obviously a pathological behavior of the model. 

\n
We conclude by noticing that the regular behavior of $h_{\underline{\alpha},\+ \theta}(R)$ for $R \to 0$ is a crucial ingredient for the study of the three-body problem with two heavy particles interacting via  contact interaction with a light particle in the Born-Oppenheimer approximation. For a detailed discussion we refer to~\cite{FST}.

\end{document}